\documentclass{elsart}
\usepackage[square]{natbib}
\usepackage{amsmath}
\usepackage{graphicx}
\journal{Nuclear Physics B}
\newcommand{\unitmatrix}{1\hspace{-0.12cm}\mbox{I}}
\begin{document}
\runauthor{Fr\'esard, Ouerdane, and Kopp}
\begin{frontmatter}
\title{Slave bosons in radial gauge: a bridge between path integral and
  hamiltonian language} 
\author[Caen]{Raymond Fr\'esard\corauthref{corauth}},
\corauth[corauth]{Corresponding author}
\ead{Raymond.Fresard@ensicaen.fr}
\author[Caen]{Henni Ouerdane\thanksref{now}},
\thanks[now]{Present address: LASMEA, UMR CNRS-Universit\'e Blaise Pascal 6602,
24 avenue des Landais, 63177 Aubi\`ere Cedex, France.}
\author[Augsburg]{Thilo Kopp}

\address[Caen]{Laboratoire CRISMAT UMR CNRS-ENSICAEN 6508, 6 Boulevard
  Mar\'echal Juin, 14050 Caen Cedex, France}
\address[Augsburg]{EP VI, Center for Electronic Correlations and Magnetism, 
  Universit\"at Augsburg, D-86135 Augsburg, Germany}

\begin{abstract}
We establish a correspondence between the resummation of world lines and the
diagonalization of the Hamiltonian for a strongly correlated electronic
system. For this purpose, we analyze the functional integrals for the
partition function and the correlation functions invoking a slave boson
representation in the radial gauge. We show in the spinless case that the
Green's function of the physical electron and the projected Green's function
of the pseudofermion coincide. Correlation and Green's functions in the
spinful case involve a complex entanglement of the world lines which, however,
can be obtained through a strikingly simple extension of the spinless
scheme. As a toy model we investigate the two-site cluster of the 
single impurity Anderson model which yields analytical results. All
expectation values and dynamical correlation functions are obtained from the
exact calculation of the relevant functional integrals. The hole density, the
hole auto-correlation function and the Green's function are computed, and a
comparison between spinless and spin 1/2 systems provides insight into the
role of the radial slave boson field. In particular, the exact expectation
value of the radial slave boson field is finite in both cases, and it is not
related to a Bose condensate.
\end{abstract}

\end{frontmatter}

\section{Introduction}\label{Sec:int}

In contemporary solid state research, strongly correlated electrons comprise
the most fascinating albeit intangible physical systems. They cover a wide
range of phenomena, including high temperature superconductivity, colossal
magnetoresistance, aspects of the fractional quantum Hall effect, and even
electronic reconstruction in oxide electronic devices which are built on
interfaces of strongly correlated films. Whereas their importance is generally
perceived, a fundamental  comprehension is still not achieved, especially for
high-temperature superconductivity. 

This unfortunate absence of a well established theoretical scheme or even
solution is not surprising: strong electronic correlations are based on
(sufficiently strong) local interactions in real space but the Fermi surface,
the concept on which the physics of metals is firmly rooted, is defined and
understood in momentum space. Correspondingly, a theoretical investigation is
either built on momentum or real space approaches which allow to treat either
the kinetic (band) term or the interaction accurately. However, a
momentum-space weak coupling approach is insufficient to generate the desired
new energy scales and fix points whereas standard perturbation theory from the
highly degenerate local (atomic) limit suffers from severe
drawbacks~\cite{Pai00}.  

Nevertheless, many effective strong coupling theories expand, in a generalized
sense, with respect to local models. A model with a single local interaction
term is the Anderson impurity model. It is the prominent strong-coupling
many-body model which can still be solved exactly (with certain restrictions)
and  which has been understood in basically all aspects
(see, for example \cite{Anderson70,Nozieres74,Wilson75,Andrei80, Wiegmann80}). 
It may justly be seen as the paradigm of a strongly correlated many-body
system. A successful scheme to investigate lattice models with on-site
interactions originates from a self-consistent extension of the Anderson
model. The self-consistency is generated through a dynamical mean-field theory
(DMFT) which singles out a site with strong local interaction; this site
couples to an electronic bath, the effective medium, the local density of
states of which is calculated
self-consistently~\cite{Georges92,Georges96}. Actually, the DMFT is exact for
infinite 
space dimensions, a limit which was introduced in Ref.~\cite{Metzner89} for 
correlated electron systems. However, it is missing the spatial correlations. 
In recent years it has been
devised to treat clusters which can couple to various bath systems in order to
investigate correlations with a spatial extension of the cluster
size~\cite{Maier05,cdmft,bk,Potthoff05,Kyung06}.  

In our theoretical study we will focus on a different approach, the slave
boson technique~\cite{BAR76,Col84,KOT86}. The formalism entails a local decomposition of
electronic excitations into charge and spin components. Electron creation and
annihilation operators are thereby represented by composite operators which
separate into canonical operators with bosonic and fermionic
character. However these operators are enslaved in the sense that their
respective number operators have to
fulfill a local constraint. The original idea was to decouple spin and charge
degrees of freedom; other, modified schemes attribute to each type of
excitation a bosonic mode which allows to study the correlated system in a
saddle point approximation~\cite{KOT86,FRE92,Has97}. This mean field approach
has been successful when set against numerical simulations:
ground state energies~\cite{Fre91} and charge structure factors show excellent
agreement~\cite{Zimmermann97},
as the procedure is exact in the large degeneracy limit~\cite{FRE92,Flo02}. 
 
While the saddle point approximation allows to calculate translationally
invariant expectation values in momentum space, the corresponding mean field
solution is not a priori legitimate. The objection is concerned with the local
decomposition of the electron field into fermion and slave boson
components. This implies that the model acquires a local gauge invariance with
the consequence that Elitzur's theorem~\cite{Elitzur75} prevents the (slave)
bosonic fields to 
acquire a non-zero expectation value. In fact, it is the phase fluctuations of
the boson field which suppress the finite value or condensation of these
fields.

One alternative to avoid such a condensation has been devised by 
Kroha, W\"olfle, and coworkers~\cite{Kroha97,Kirchner04}. In their approach the
local gauge invariance is 
guaranteed through Ward identities in a conserving approximation. The
projection onto the physical sector of the Fock space is achieved with an
Abrikosov procedure by  sending the Lagrange multiplier of the constraint to
infinity.\footnote{Note that a similar procedure can be set up without
introducing slave bosons~\cite{Baumgartner05}.} 
The other alternative is to use a radial decomposition of the bosonic field
\cite{Read}, the details of which were presented in a previous paper by two of
the authors~\cite{FRE01}. In the limit of large on-site 
interaction, the bosonic fields in radial representation reduce to their
respective (real) amplitude as the time derivatives of the conjugate phase can
be absorbed in a time-dependent Lagrange multiplier field. 

In this article we provide a scheme for the solution of cluster models in
radial slave-boson representation. We present in sufficient detail the
calculation of correlation and Green's functions for a two-site cluster of the
single impurity Anderson model, in order to exemplify our scheme. Although the
model can 
be diagonalized without slave boson technique we esteem the explicit solution
in the radial decomposition of considerable significance. First, it relates
the world line expansion of slave boson path integrals to the quantum states
in the Fock space, in particular for entangled states. This is achieved
through a 
decomposition of the fermionic determinant into resolvents at each time
step. Second, it allows to compare these exact results (e.~g., for the
slave boson amplitude) to saddle point evaluations and to
assess their validity~\cite{OUE07}.  

The article is organized as follows: in Sec.~\ref{Sec:FI}, we introduce the
functional integral formulation of the two-site cluster model. We give
expressions for the action, partition function and hole density as well as for
the hole autocorrelation function in terms of radial slave bosons. The
spinless system is studied first in Sec.~\ref{Sec:spinless} where, through the
derivation of the partition function, we show how to proceed from a world line
representation to the representation with quantum states in the Fock space. In
Sec.~\ref{Sec:spinful}, we show how our formalism allows to derive results for
the spinful case from a straightforward extension of the spinless case. 
The Green's function necessitates a distinct derivation of the fundamental
connection between the slave-boson path integral representation and the
hamiltonian scheme which is the object of Sec.~\ref{Sec:GF}. 

\section{Functional integral formulation of the two-site cluster
  model}\label{Sec:FI} 

\subsection{Hamiltonian and radial slave boson representation}

The single impurity Anderson model (SIAM) has been investigated with a variety
of techniques and for many different purposes. One of them consists of testing
a new approach, in particular against exact results. Here we adopt a similar
spirit in order to link the evaluation of the path integral representation of
thermodynamic and dynamical quantities to their computation through a
straightforward diagonalization of the Hamiltonian. For the SIAM it reads:  

\begin{equation}\label{eqh00}
{\mathcal H} = \sum_{k,\sigma} c^{\dagger}_{k,\sigma}
(t^{\phantom{\dagger}}_k+\epsilon_{\rm c})
c^{\phantom{\dagger}}_{k,\sigma} + \sum_{\sigma} d^{\dagger}_{\sigma}
\epsilon_{\rm d}  d^{\phantom{\dagger}}_{\sigma} + V
\sum_{k,\sigma}
\left(c^{\dagger}_{k,\sigma}d^{\phantom{\dagger}}_{\sigma} + {\rm
  h.c.}\right)+ U \prod_{\sigma = \uparrow,\downarrow} d^{\dagger}_{\sigma}
d^{\phantom{\dagger}}_{\sigma} ,
\end{equation}

\noindent where $U$ is the  on-site repulsion, which is hereafter taken as
infinite. The operators $c^{\dagger}_{k,\sigma}$
($c^{\phantom{\dagger}}_{k,\sigma}$) and $d^{\dagger}_{\sigma}$
($d^{\phantom{\dagger}}_{\sigma}$) describe the creation (annihilation) of the
band electrons and impurity electrons respectively, with spin $\sigma$; the
kinetic energy in the band is denoted $t^{\phantom{\dagger}}_k$,  
and $\epsilon_{\rm  c}$ and $\epsilon_{\rm d}$ are the band and impurity 
energy levels, respectively.
The hybridization is given by $V$.  

The link between the two schemes is actually a complex procedure when the
impurity is coupled to an infinite bath. In order to lower this complexity we
reduce the bath to a single site, in which case the diagonalization of
the Hamiltonian is easy and all relevant results can be obtained
analytically. Nevertheless, the problem is non-trivial when handled in the
functional integral formalism. The level of difficulty depends on the
functional representation which is used. For our purpose, a promising one is
that of the slave boson (SB) representation in the radial gauge
\cite{FRE01}. It is based on the original representation by Barnes
\cite{BAR76} and is augmented in that respect that the underlying $U(1)$
gauge symmetry, originally discussed by Read and Newns \cite{Read}, is fully
implemented, as the phase of the bosonic field is integrated out from the outset. 
Accordingly, the original field $d^{\phantom{\dagger}}_{\sigma}$ is represented in 
terms of a real and a Grassmann field for each spin:
    
\begin{eqnarray}
\label{eqh01a}
d^{\phantom{\dagger}}_{n,\sigma} & = & x^{\phantom{\dagger}}_{n+1}
f^{\phantom{\dagger}}_{n,\sigma}\\ 
\label{eqh01b}
d^\dagger_{n,\sigma} & = & x^{\phantom{\dagger}}_n f^\dagger_{n,\sigma},
\end{eqnarray}

\noindent where $x^{\phantom{\dagger}}_n $ and
$x^{\phantom{\dagger}}_{n+1}$ are the slave boson field amplitudes at time
steps $n$ and $n+1$, and $f^{\phantom{\dagger}}_{n,\sigma}$ is the auxiliary
fermion field. The shift of one time step for $x$ in the relation for
$d^{\phantom{\dagger}}_{n,\sigma}$ is necessary to obtain a non-zero value of
the Grassmann integration for the Green's functions $-\langle 
d^{\phantom{\dagger}}_{\sigma} (\tau)d^{{\dagger}}_{\sigma} (0)\rangle $  as
clearly shown for its calculation in the atomic limit in
Ref.~\cite{FRE01}. More precisely, the path integral is zero if
$x^{\phantom{\dagger}}_{n+1}$ is replaced by $x^{\phantom{\dagger}}_n$ in
Eq.~(\ref{eqh01a}). Moreover, Eqs.~(\ref{eqh01a}) and (\ref{eqh01b}) are  
required in order to properly represent the hybridization term in the action
as given below. Further detail on this matter can be found in
Ref.~\cite{FRE01}.

\subsection{Action and partition function}

Following Ref.~\cite{FRE01}, the path integral representation of the partition
function of the two-site cluster is given by: 

\begin{equation}\label{eqh0}
{\mathcal Z} = \lim_{\substack{N\rightarrow\infty \\W\rightarrow\infty}}
     \left(\prod_{n=1}^N \int \prod_{\sigma}
D[f_{n,\sigma},f^\dagger_{n,\sigma}]
D[c_{n,\sigma},c^\dagger_{n,\sigma}] \int_{-\infty}^{\infty}
\frac{\displaystyle \delta {\rm d}\lambda_n}{\displaystyle 2\pi}
\int_{-\infty}^{\infty} {\rm d}x_n\right) e^{-S}
\end{equation}

\noindent where the action $S$ may be written as the sum of a fermionic part,
     $S_f$, and a bosonic part, $S_b$, with 

\begin{eqnarray}\label{eqh1}
S_f & = & \sum_{\sigma} S_{f,\sigma} = \sum_{n=1}^N \sum_{\sigma}
  \left[c^\dagger_{n,\sigma}(c_{n,\sigma} - 
  L_{\rm c} c_{n-1,\sigma})+f^\dagger_{n,\sigma}(f_{n,\sigma} - L_{
    n} f_{n-1,\sigma})\right.\\
    \nonumber 
    && ~~~~~~~~~~~~~~~~~~~~~~~\left. + V\delta x_n(c^\dagger_{n,\sigma}
  f_{n-1,\sigma} 
  + f^\dagger_{n,\sigma}c_{n-1,\sigma})\right]\\
  \nonumber 
S_b & = &\sum_{n=1}^N
  \left[\delta\left(i\lambda_n(x_n-1)+Wx_n(x_n-1)\right)\right], 
\end{eqnarray} 

\noindent where $L_{\rm c} = e^{-\delta(\epsilon_{\rm c} - \mu)}$, $L_{ n} =
e^{-\delta(\epsilon_{\rm d} - \mu + i\lambda_n)} \equiv L_{\rm
d}~e^{-i\delta\lambda_n}$, $\lambda_n$ is the time-dependent real constraint
field, $n$ denotes the time steps, and $\delta \equiv \beta/N$, with $\beta
= 1/k_{\rm B}T$ and $N$ the number of time steps. Here, $S_f$
($S_{f,\sigma}$) is bilinear in the fermionic fields, and the corresponding
matrix of the coefficients will be denoted as $\left[S\right]$
($\left[S_0\right]$).  
The positive real number $W$ is sent to infinity at the end of the calculation
which guarantees the projection onto the physical subspace. 
The above treatment of the bosonic field is specific to radial slave bosons:
no phase variable appears, and the above form cannot be obtained by
transformations of the conventional functional integral in the Cartesian gauge
without further assumptions (see Ref.~\cite{FRE01}). 

Inserting Eq.~(\ref{eqh1}) into Eq.~(\ref{eqh0}) yields a particularly
suggestive expression of the partition function ${\mathcal Z}$ in the
functional integral formulation, of the kind:   

\begin{equation}\label{eqh2}
{\mathcal Z} = \lim_{\substack{N\rightarrow\infty \\ W\rightarrow\infty}}
     {\mathcal P_1}\ldots{\mathcal P_N}~\mbox{det}
\left[S\right], 
\end{equation}

\noindent where det$\left[S\right]$ is the determinant of the fermionic matrix
     defined by Eq.~(\ref{eqh1}); ${\mathcal P}_n$ is defined as:  

\begin{equation}\label{eqh3}
{\mathcal P}_n = \int_{-\infty}^{+\infty} \delta~\frac{\displaystyle
  {\rm d}\lambda_n}{\displaystyle 2\pi}~ \int_{-\infty}^{+\infty} {\rm
  d}x_n~e^{-\delta\left[i\lambda_n(x_n-1)+Wx_n(x_n-1)\right]}, 
\end{equation}

\noindent and acts as a projector from the enlarged Fock space ``spanned''
  by the 
auxiliary fermionic fields down to the physical one. Explicitly, the action of
these projectors on the various contributions resulting from
$\mbox{det}\left[S\right]$ are found to be:  

\begin{eqnarray}
\label{eqh4a}{\mathcal P}_n \cdot 1 & = & 1,\\
\label{eqh4b}{\mathcal P}_n \cdot x_n & = & 1,\\
\label{eqh4d}{\mathcal P}_n \cdot L_{\rm n} & = & L_d,\\
\label{eqh4e}{\mathcal P}_n \cdot L_{n}x_n & = & 0,\\
\label{eqh4f}{\mathcal P}_n \cdot L_{n}^2 & = & 0,\\
\label{eqh4g}{\mathcal P}_n \cdot x_n^2 & = & 1.
\end{eqnarray}

As will be seen below no further property of ${\mathcal P}_n$ will be needed
for our purpose.  
We note that there is some freedom in writing the projectors ${\mathcal P}_n$,
and alternative expressions exist \cite{FRE01}. However the properties
Eqs.~(\ref{eqh4a})--(\ref{eqh4g}) are independent of the particular form of
${\mathcal P}_n$. 

\section{Application to the spinless fermion case}\label{Sec:spinless}

We first consider a spinless fermion system for simplicity. Even though this
is a non-interacting problem, the level of complexity of its path integral
representation following from Eqs.~(\ref{eqh0})--(\ref{eqh2}), is equivalent
to the one of a fully interacting problem. The matrix representation of the
action $S_{f,\sigma}$ of such system is a $2N\times 2N$ square matrix whose
explicit expression in the basis $\{c_{n,\sigma},f_{n,\sigma}\}$ reads:  

\begin{eqnarray}\label{eqh6}
\left[S_0\right] = 
\left( \begin{array}{cccc}
\unitmatrix_2&~&~&-\left[{\mathcal L}_1\right]\\
\left[{\mathcal L}_2\right]& \unitmatrix_2 &~&~\\
~&\ddots&\ddots&~\\
~&~&\left[{\mathcal L}_{\rm N}\right]&\unitmatrix_2
\end{array}\right),
\end{eqnarray}

\noindent where $\unitmatrix_2$ is the 2$\times$2 identity matrix and
$\left[{\mathcal L}_n\right]$ are 2$\times$2 blocks given by:  

\begin{eqnarray}\label{eqh7}
\left[{\mathcal L}_n\right] = 
\left( \begin{array}{cc}
-L_{\rm c} &\delta Vx_n\\
\delta Vx_n & -L_{n}
\end{array}\right),
\end{eqnarray}

\noindent at time step $n$, $1\leq n \leq N$. Note that the matrix
$\left[S_0\right]$ as defined in Eq.~(\ref{eqh6}) has the same structure as
the action matrix $S^{(\alpha)}$ in Chap.~2 of Ref.~\cite{NEG88}. 

\subsection{Partition function}

The partition function ${\mathcal Z}_0$ of the spinless fermion system has a
form similar to that of Eq.~(\ref{eqh2}) except that
$\mbox{det}\left[S\right]$ is replaced by $\mbox{det}\left[S_0\right]$. Its
calculation is straightforward since we only have to evaluate:  

\begin{equation}\label{eqh7b}
{\mathcal Z}_0 = \lim_{\substack{N\rightarrow\infty \\ W\rightarrow\infty}}
     {\mathcal P_1}\ldots{\mathcal
  P_N}~\sum_{\{P\}} \mbox{sgn}(P) \prod_{n=1}^{2N} S_{P(n),n} 
\end{equation}

\noindent where $\mbox{sgn}(P)$ is the signum function of permutations $P$ 
in the permutation group ${\mathcal S}_{2N}$ and $S_{i,j}$ are the matrix elements of
$\left[S_0\right]$. Since the time step $n$ is only involved in
$S_{i,2n-3}$ and $S_{i,2n-2}$ we may recast Eq.~(\ref{eqh7b}) 
into:

\begin{eqnarray}\label{eqh7c}
{\mathcal Z}_0 &=& \lim_{\substack{N\rightarrow\infty \\ W\rightarrow\infty}}
  \sum_{\{P\}} \mbox{sgn}(P)~{\mathcal 
  P}_2\left(S_{P(1),1}S_{P(2),2}\right) \ldots \nonumber\\
&\times & {\mathcal 
  P}_N\left(S_{P(2N-3),2N-3}S_{P(2N-2),2N-2}\right) {\mathcal P}_1
  \left(S_{P(2N-1),2N-1}S_{P(2N),2N}\right). 
\end{eqnarray}

At this point it is straightforward to verify that performing the projections
only implies to make use of Eqs.~(\ref{eqh4a})--(\ref{eqh4d}). 
We are left with: 

\begin{equation}\label{eqh7d}
{\mathcal Z}_0 = \lim_{N\rightarrow\infty} ~\sum_{\{P\}}\mbox{sgn}(P)
\prod_{n=1}^{2N} S_{P(n),n}' = \lim_{N\rightarrow\infty} ~\mbox{det}
\left[S_0'\right], 
\end{equation}

\noindent where $S_{i,j}'$ are the elements of the $2 N\times 2 N$ matrix
$\left[S_0'\right]$ defined as:  

\begin{eqnarray}\label{eqh7e}
\left[S_0'\right] = 
\left( \begin{array}{cccc}
\unitmatrix_2&~&~&-\left[{\mathcal L}\right]\\
\left[{\mathcal L}\right]& \unitmatrix_2 &~&~\\
~&\ddots&\ddots&~\\
~&~&\left[{\mathcal L}\right]&~~\unitmatrix_2
\end{array}\right).
\end{eqnarray}

\noindent In Eq.~(\ref{eqh7e}) the 2$\times$2 matrix blocks 
$\left[{\mathcal L}\right]$
are similar to the blocks $\left[{\mathcal L}_n\right]$ except that $L_{n}$
becomes $L_{\rm d}$, and $x_n$ is replaced by 1: 

\begin{eqnarray}\label{eqh7f}
\left[{\mathcal L}\right] = 
\left( \begin{array}{cc}
-L_{\rm c} &\delta V\\
\delta V & -L_{\rm d}
\end{array}\right).
\end{eqnarray}

In the form of Eq.~(\ref{eqh7d}) it is now obvious that ${\mathcal Z}_0$ can
be readily obtained.  
We notice that Eq.~(\ref{eqh7e}) is the expected action matrix
for this free fermionic problem. 
Besides, it is straightforward to extend the above calculation to the case of
an arbitrary bath. Unfortunately the calculation becomes considerably more
involved in the spin 1/2 case, which leads us to develop another strategy to
that purpose. We first present it in the spinless case, before extending it to
the spinful case. 

\subsubsection{Generation and resummation of the world lines}

Part of the difficulty in computing ${\mathcal Z}_0$ is that the time steps
are mixed in the fermionic determinant, in contrast to the bosonic part of the
action represented by the projectors ${\mathcal P}_n$,
Eq.~(\ref{eqh2}). Therefore, transforming this determinant into a form where
the time steps are decoupled, is desirable. Achieving this amounts to handle
all the world lines following from the action in Eq.~(\ref{eqh1}) which
represents a problem equivalent to a particle in a time-dependent field with a
time-dependent hopping amplitude. 

In order to generate the dynamics of the world lines we first expand
$\mbox{det}\left[S_0\right]$ along the first two columns. 
We obtain:

\begin{eqnarray}\label{eqh8}
\mbox{det}\left[S_0\right] & = & 1\times M_{1,2}^{\overline{2}} +
L_{\rm c} M_{3,2}^{\overline{2}} + \delta Vx_2
M_{4,2}^{\overline{2}}- \delta Vx_2 M_{1,3}^{\overline{2}}- L_2
M_{1,4}^{\overline{2}} \\ 
\nonumber
&&+ \left(L_{\rm c}L_2 - (\delta V x_2)^2 \right) M_{3,4}^{\overline{2}}. 
\end{eqnarray}

\noindent Here the notation is as follows: we construct a matrix similar to
$\left[S_0\right]$, but we only include time steps 1 and $m >
n$. $M_{i,j}^{\overline{n}}$ is a minor of this matrix, where both the $i$-th
and $j$-th rows, together with the first and second columns, are removed.

At this stage we may proceed with the generation of the world lines. To that
aim we need to express 
the minors $M^{\overline{2}}$ as linear combinations of
the minors $M^{\overline{3}}$ which in turn can also be expressed as similar
linear combinations of the minors $M^{\overline{4}}$, and so forth up to the
time step $N$. The recurrence relation that we have established takes the
following form:  

\begin{eqnarray}\label{eqh9}
\left(\begin{array}{c}
M^{\overline{n}}_{1,2}\\
M^{\overline{n}}_{3,2}\\
M^{\overline{n}}_{4,2}\\
M^{\overline{n}}_{1,3}\\
M^{\overline{n}}_{1,4}\\
M^{\overline{n}}_{3,4}
\end{array}\right)=
\left(\begin{array}{cccc}
\left[K^{(n+1),1}\right]&~&~&~\\
~&\left[K^{(n+1),2}\right]&~&~\\
~&~&\left[K^{(n+1),3}\right]&~\\
~&~&~&\left[K^{(n+1),4}\right]
\end{array}\right)
\left(\begin{array}{c}
M^{\overline{n+1}}_{1,2}\\
M^{\overline{n+1}}_{3,2}\\
M^{\overline{n+1}}_{4,2}\\
M^{\overline{n+1}}_{1,3}\\
M^{\overline{n+1}}_{1,4}\\
M^{\overline{n+1}}_{3,4}
\end{array}\right),
\end{eqnarray}

\noindent where the four matrix blocks

\begin{eqnarray}
\label{eqh9a}\left[K^{(n),1}\right] & = & (1),\\
\label{eqh9b}\left[K^{(n),2}\right] & = & \left[K^{(n),3}\right]=
\left(\begin{array}{cc} 
L_{\rm c}&\delta Vx_n\\ \delta Vx_n& L_{n}
\end{array}\right),\\
\label{eqh9c}\left[K^{(n),4}\right] & = & (L_{\rm c}L_{n}),
\end{eqnarray}

\noindent at time step $n$, $1\leq n\leq N$, define the 6$\times$6 block
diagonal matrix $\left[K^{(n)}\right]$. \footnote{In Eq.~(\ref{eqh9c}) a term
  of order $\delta^2$, that vanishes in the limit $N\rightarrow \infty$, is
  neglected.}
The matrices $\left[K^{(n)}\right]$
describe the evolution of the two-site system along the world lines at each
time step $n$. Iterating this procedure up to time step $N$ yields the
determinant det$\left[S_0\right]$ in the following scalar product form:  

\begin{equation}\label{eqh10}
\mbox{det}\left[S_0\right] = 
\left(\begin{array}{cccccc}
1,&L_{\rm c},&\delta Vx_2,&-\delta Vx_2,&-L_2,&L_{\rm c}L_2
\end{array}\right)
~\prod_{n=3}^N \left[K^{(n)}\right]
\left(\begin{array}{c}
1\\
L_{\rm c}\\
\delta Vx_1\\
-\delta Vx_1\\
-L_1\\
L_{\rm c}L_1
\end{array}\right), 
\end{equation}

\noindent where the row vector is identified from Eq.~(\ref{eqh8}) and the column
vector has been obtained from the last time step of the iteration process.
Since the minus signs in Eq.~(\ref{eqh10}) cancel, they can be discarded for further
considerations. 

\begin {figure}[!t]
$ $\\[1em]
\centering
\scalebox{.45}{\rotatebox{0}{\includegraphics*{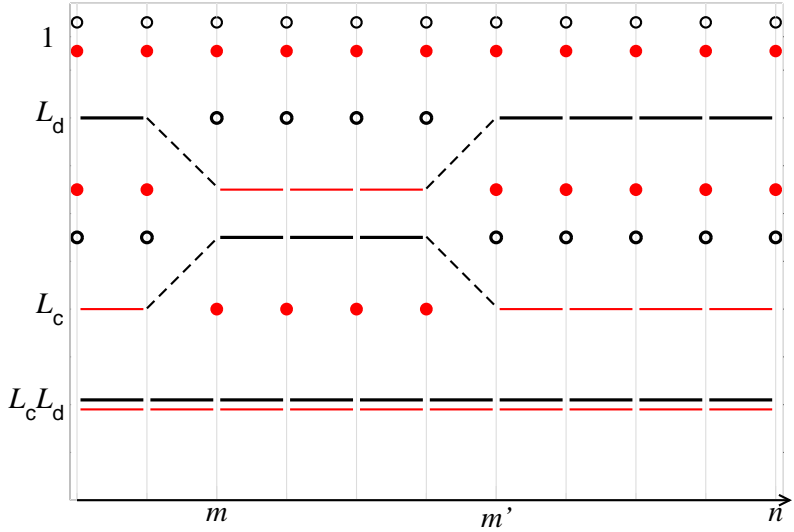}}}
\caption{Typical world lines of a two-site system. Thick (thin) lines denote
  the elementary processes with the electron sitting on the impurity (``band'')
  site. Full (empty) circles denote
  the elementary processes with a hole sitting on the ``band'' (impurity)
  site. Dashed lines represent hopping processes. The horizontal axis runs
  with the time steps, while in the vertical direction different sites are
  displayed.} 
\label{fig1} 
\end{figure}

The above expression corresponds to the full resummation of the world lines,
some of them being represented in Fig.~\ref{fig1}. The first contribution,
labeled by $1$, corresponds to the
subspace with zero electron while the last one, labeled by $L_{\rm c}L_{\rm
  d}$, 
corresponds to the subspace 
with two electrons. In both cases the world lines are ``straight'': namely no
hopping process takes place, and the system is right away in an eigenstate.
In Eq.~(\ref{eqh10}) they correspond to the terms involving
$\left[K^{(n),1}\right]$  and $\left[K^{(n),4}\right]$ respectively. The
structure of the world lines in the one-electron subspace is more intricate.  

In order to gain an intuitive picture of these world lines let us first
consider the 
trivial functional integral representation of an interactionless electron
where the $f$-field directly represents the physical
electron. 
We begin with the processes where the electron 
is on the ``band'' site at time step one. If it stays there during all time
steps, the resulting contributions ${\mathcal Z}_{\rm c}^{(0)}$ to ${\mathcal
  Z}_0$ will be given by ${\mathcal Z}_{\rm c}^{(0)} = \prod_{n=1}^N L_{\rm
  c}$, 
namely there is one factor $L_{\rm c}$ per time step and the world line is
straight. If, on the contrary, the electron hops onto the impurity at time step
$m$, and back to the ``band'' at time step $m'$, the corresponding contribution
${\mathcal Z}_{mm',{\rm c}}^{(2)}$ to ${\mathcal Z}_0$ results in ${\mathcal
  Z}_{mm',{\rm c}}^{(2)}=\left(\prod_{n=1}^{m-1} L_{\rm c}\right)\delta
V\left(\prod_{n=m+1}^{m'-1}L_{\rm d}\right) 
\delta V\left(\prod_{n=m'+1}^{N} L_{\rm c}\right)$. 
Higher order processes in $V$ follow accordingly. Complementary processes
are those where the electron resides on the impurity at
time step one. The
contributions of all these processes to ${\mathcal Z}_0$ will be weighted by
both the number of hopping processes and the difference in energy between the
two levels. Assuming $\epsilon_d<\epsilon_c$ results in $L_{\rm c}<L_{\rm d}$,
and for 
world lines involving the same number of hopping process, the world line
containing the largest number of factors $L_{\rm d}$ yields the largest
contribution.

If we now return to our representation the contributions of the world lines to
the partition function follow in a similar fashion, except for that i) the
factor corresponding to hopping process at time step $m$ is given by $\delta V
x_m$, and ii) the particle line at time step $n$ corresponding to
the electron sitting on the impurity site results in a factor
$L_{n}$. Accordingly the contribution of the world line labeled by
$L_{\rm c}$ in 
Fig.~\ref{fig1} to $\mbox{det}\left[S_0\right]$ is:
\begin{equation}
\left(\prod_{n=1}^{m-1} L_{\rm c}\right)
\delta V x_m \left(\prod_{n=m+1}^{m'-1}L_{n}\right) 
\delta V x_{m'} \left(\prod_{n=m'+1}^{N} L_{\rm c}\right), 
\end{equation}
while the world
line labeled by $L_{\rm d}$ in Fig.~\ref{fig1} yields 
\begin{equation}\left(\prod_{n=1}^{m-1}
  L_{n}\right) 
\delta V x_m \left(\prod_{n=m+1}^{m'-1}L_{\rm c}\right) 
\delta V x_{m'} \left(\prod_{n=m'+1}^{N} L_{ n}\right).
\end{equation}

\subsubsection{Connecting with the hamiltonian language}

We observe that the structure of Eq.~(\ref{eqh10}) is not manifestly translationally
invariant in time. For this reason, we proceed to bring  Eq.~(\ref{eqh10}) into a form
where no particular time step is singled out. 
We  note that the (row) column vector in Eq.~(\ref{eqh10}) can be identified
with the (rows) columns of the matrix blocks 
$\left[K^{(n)}\right]$ --- disregarding the cancelling minus signs.
Accordingly, we can rewrite Eq.~(\ref{eqh10}) as: 

\begin{eqnarray}\label{eqh12}\nonumber
\mbox{det}\left[S_0\right] && =
\sum_{\{\alpha_2,\ldots,\alpha_N\}}~K^{(2),1}_{1\alpha_2}~ K^{(3),1}_{\alpha_2
  \alpha_3}\ldots K^{(N),1}_{\alpha_{N-1} \alpha_N} K^{(1),1}_{\alpha_N 1}\\ 
\nonumber
&& + \sum_{\{\alpha_2,\ldots,\alpha_N\}}~K^{(2),2}_{1\alpha_2}~
K^{(3),2}_{\alpha_2 \alpha_3}\ldots K^{(N),2}_{\alpha_{N-1} \alpha_N}
K^{(1),2}_{\alpha_N 1}\\ 
\nonumber
&& + \sum_{\{\alpha_2,\ldots,\alpha_N\}}~K^{(2),3}_{2\alpha_2}~
K^{(3),3}_{\alpha_2 \alpha_3}\ldots K^{(N),3}_{\alpha_{N-1} \alpha_N}
K^{(1),3}_{\alpha_N 2}\\ 
&& + \sum_{\{\alpha_2,\ldots,\alpha_N\}}~K^{(2),4}_{1\alpha_2}~
K^{(3),4}_{\alpha_2 \alpha_3}\ldots K^{(N),4}_{\alpha_{N-1} \alpha_N}
K^{(1),4}_{\alpha_N 1}, 
\end{eqnarray}

\noindent since the matrices $\left[K^{(n)}\right]$ are block diagonal and
symmetric. Observe that the first and last lower index in line three of  Eq.~(\ref{eqh12})
is not 1 but 2.

In Eq.~(\ref{eqh12}), the first sum is equal to 1; the second and third sums
are the diagonal elements of the matrix product $\prod_{n=1}^N
\left(\begin{array}{cc} L_{\rm c}& \delta Vx_n \\ \delta Vx_n & L_{n}
  \end{array}\right)$, respectively. The last sum is equal to $L_{\rm c}^N
\prod_{n=1}^N L_{n}$. Therefore Eq.~(\ref{eqh12}) reduces to the trace of a
4$\times$4 matrix that is the product of the $N$ block diagonal matrices
$\left[{\mathcal K}_n\right]$ whose elements are $\left[K^{(n),1}\right]$,
$\left[K^{(n),2}\right]$ and $\left[K^{(n),4}\right]$, respectively:  

\begin{eqnarray}\label{eqh13}
\mbox{det}\left[S_0\right] = \mbox{Tr} \prod_{n=1}^N
\left[{\mathcal K}_n\right],
\end{eqnarray}

\noindent where the matrix $\left[{\mathcal K}_n\right]$ is given by:

\begin{equation}\label{eqK}
\left[{\mathcal K}_n\right]=
\left(\begin{array}{cccc}
1&~&~&~\\
~&L_{\rm c}&\delta Vx_n&~\\
~&\delta Vx_n&L_{n}&~\\
~&~&~&L_{\rm c}L_{n}
\end{array}\right).
\end{equation}

Here, it is of interest to note that we have established a direct link between
the world line picture embodied in the 6$\times$6 matrices
$\left[K^{(n)}\right]$ and the simpler description in terms of quantum states
in the Fock space through the 4$\times$4 matrices $\left[{\mathcal
    K}_n\right]$. Indeed, when performing the world line expansion using
Eq.~(\ref{eqh10}), the propagation matrices $\left[K^{(n)}\right]$ acquire an
involved structure following from the initial conditions attached to the world
lines. This is best seen in the one-electron sub-space: at time-step one, the
electron may either be on site {\rm c} or on site {\rm d} (see
Fig.~\ref{fig1}), and the corresponding dynamics is governed by the matrices
$\left[K^{(n),2}\right]$ and $\left[K^{(n),3}\right]$. Summing up all these
world lines yields the one-electron contribution to $\mbox{det}
\left[S_0\right]$. Thus, one needs to handle two 2$\times$2 matrices, one in
each case. In contrast, a single 2$\times$2 matrix needs to be treated on the
level of Eq.~(\ref{eqh13}).  

In the above form of the fermionic determinant, Eq.~(\ref{eqh13}), the time
steps are decoupled, which greatly simplifies the projection onto the physical
Hilbert space. Indeed, in terms of the matrix $\left[q\right]$  given by: 

\begin{equation}\label{eqq}
\left[q\right] \equiv {\mathcal P_n}\left(\left[{\mathcal K}_n\right]\right) = 
\left(\begin{array}{cccc}
1&~&~&~\\
~&L_{\rm c}&\delta V&~\\
~&\delta V&L_{\rm d}&~\\
~&~&~&L_{\rm c}L_{\rm d}
\end{array}\right), 
\end{equation}

we obtain the partition function ${\mathcal Z}_0$ as: 

\begin{eqnarray}\label{eqh13b}
{\mathcal Z}_0 = \lim_{N\rightarrow\infty}\mbox{Tr} \prod_{n=1}^N
\left[ q \right]
=\!\!\lim_{N\rightarrow\infty}\!\!\mbox{Tr}
\left[\unitmatrix_4 \!\!-\! \delta
\left(\begin{array}{cccc}
0&~&~&~\\
~&\epsilon_{\rm c}-\mu&-V&~\\
~&-V&\epsilon_{\rm d} - \mu&~\\
~&~&~&\epsilon_{\rm c}+\epsilon_{\rm d} - 2\mu
\end{array}\right)\right]^N
\end{eqnarray}

Namely, we recover here the hamiltonian matrix in the Fock space. Therefore
the expansion of $\mbox{det} \left[S_0\right]$ in minors together with the
recurrence relations, Eq.~(\ref{eqh9}), allows for a correspondence between the
ensemble of the world lines and the hamiltonian matrix. It is apparent that
the complexity of this interrelation depends on whether or not the system is in
an eigenstate at time step one. If this is the case, there is one single
straight world line, and the connection is obvious. Otherwise there is a
proliferation of world lines, here controlled by the matrices
$\left[K^{(n),2}\right]$ and  $\left[K^{(n),3}\right]$, which recombine to
yield the contribution resulting from the 2$\times$2 block of the matrix
$\left[{\mathcal K}_n\right]$. A result equivalent to Eq.~(\ref{eqh13b}) was
already obtained by Barnes \cite{BAR77}, though in a totally different
fashion. 

\subsection{Hole density}

In our path integral formalism,
the expectation value of the amplitude of the slave boson field at time step
$m$ is $\langle x_m \rangle$. It simply represents the hole density $1-n_{\rm d}$
which can be written as
$\langle x_m \rangle = \langle b^{\dagger}_m b^{\phantom{\dagger}}_{m-1}
\rangle$ where $b$ represents the original Barnes slave boson, and $\langle x_m \rangle$
is finite. In contrast the expectation values of the boson operators are zero:
$\langle b^{\dagger}_m\rangle = \langle b_m \rangle = 0$, for each time step
because of the fluctuations of their respective phase factor, in agreement with
Elitzur's theorem. Note that expectation values of higher order moments of $x_m$
are also non zero: $\langle x^a_m \rangle = \langle x_m \rangle \neq 0$ for
any real positive parameter $a$. 

In the case of spinless fermions we calculate $\langle x_m \rangle$ as:

\begin{eqnarray}\label{eqhx1}
{\mathcal Z}_0\langle x_m\rangle & = & \lim_{\substack{N\rightarrow\infty \\
  W\rightarrow\infty}} {\mathcal 
  P_1}\ldots{\mathcal P_N}~\left(\mbox{det}
\left[S_0\right]~x_m\right)\\
\nonumber
& = &\lim_{\substack{N\rightarrow\infty \\ W\rightarrow\infty}} {\mathcal
  P_1}\ldots{\mathcal P_N}~\left(x_m \mbox{Tr}~\prod_{n=1}^N
\left[{\mathcal K}_n\right]\right) .
\end{eqnarray}

If we introduce the 4$\times$4 matrix
$\left[{\mathcal Q}_X\right] \equiv {\mathcal P}_n (x_n
\left[{\mathcal K}_n\right])$ for all $n$,  and $\left[{\mathcal Q}\right]$ 
the 4$\times$4 diagonal matrix  
that satisfies
$\left[q\right] = \left[U_{\mathcal
    Q}\right]\left[{\mathcal Q}\right]\left[{U}_{\mathcal Q}\right]^{\dagger}$,
$\left[U_{\mathcal Q}\right]$ being the eigenvector matrix,
Eq.~(\ref{eqhx1}) becomes: 

\begin{equation}\label{eqhx2}
{\mathcal Z}_0\langle x_m\rangle = \lim_{N\rightarrow\infty}
\mbox{Tr} \left(\left[{\mathcal Q}\right]^{N-1} \left[{U}_{\mathcal
    Q}\right]^{\dagger} \left[{\mathcal Q}_X\right]\left[U_{\mathcal
    Q}\right]\right).
\end{equation}

Using Eqs.~(\ref{eqh4a})--(\ref{eqh4d}), and the $\delta \rightarrow 0$ limit,
we obtain that the matrix $\left[{\mathcal Q}_X\right]$ reduces to the
representation of the hole density operator on the impurity in the Fock space: 

\begin{equation}\label{eqdefqx}
\left[{\mathcal Q}_X\right]_{i,j}= \delta_{i,1}~\delta_{j,1} +
\delta_{i,2}~\delta_{j,2} .
\end{equation}
Correspondingly, Eq.~(\ref{eqhx2}) expresses $\langle x_m \rangle$ as the
averaged value of the hole density operator on the impurity, represented by its
matrix elements in the basis of the eigenstates of the Hamiltonian
($\left[{U}_{\mathcal 
    Q}\right]^{\dagger} \left[{\mathcal Q}_X\right]\left[U_{\mathcal
    Q}\right]$)
and weighted by the Boltzmann factors $ \frac{1}{{\mathcal Z}_0}
\left[{\mathcal Q}\right]^{N-1}$. 
Therefore, in contrast to a Bose condensate,  $\langle x_m \rangle$ is
generically finite and may only vanish for zero hole concentration.
Its numerical evaluation will be presented in a forthcoming
paper~\cite{OUE07}.

\subsection{Density-density correlation function}

To obtain further insight into the approach, it is of interest to compute
dynamical correlation functions. Probably, the simplest one is provided by the
hole density autocorrelation function on the impurity site. When expressed in
terms of eigenstates of the Hamiltonian, it takes the form:  

\begin{equation}\label{eqh17a}
\langle \left(1-n_{\rm d}(m\delta)\right)\left(1-n_{\rm
    d}(\delta)\right)\rangle = \frac{\displaystyle 1}{\displaystyle {\mathcal
    Z}_0}~\sum_{\alpha,\alpha'} e^{\delta(m -N) E_{\alpha}} e^{-m\delta
  E_{\alpha'}} |\langle 
\psi_{\alpha}|1-n_{\rm d}|\psi_{\alpha'}\rangle|^2,  
\end{equation}

\noindent where the eigenvalues $E_{\alpha}$ can be obtained from the
diagonalization of the Hamiltonian. The latter can be read from,
e.~g., Eq.~(\ref{eqh13b}).  

In our path integral representation, the autocorrelation function $\langle
x_1x_m\rangle$ may be written as:  

\begin{eqnarray}\label{eqh18}
{\mathcal Z}_0\langle x_1x_m\rangle & = & \lim_{\substack{N\rightarrow\infty
  \\ W\rightarrow\infty}} {\mathcal  
  P_1}\ldots{\mathcal P_N}~\left(\mbox{det}
\left[S_0\right]~x_1x_m\right)\\
\nonumber
& = & \lim_{\substack{N\rightarrow\infty \\ W\rightarrow\infty}} {\mathcal
  P_1}\ldots{\mathcal P_N}~\left(x_1x_m \mbox{Tr}~\prod_{n=1}^N
\left[{\mathcal K}_n\right]\right),
\end{eqnarray}

\noindent which reduces to

\begin{equation}\label{eqh19}
{\mathcal Z}_0\langle x_1x_m\rangle = \lim_{N\rightarrow\infty}
\mbox{Tr} \left(\left[{\mathcal Q}_X\right]\left[q\right]^{m-2}\left[{\mathcal
      Q}_X\right]\left[q\right]^{N-m}\right), 
\end{equation}

\noindent after application of the projectors, 
Eqs.~(\ref{eqh4a})--(\ref{eqh4g}). 
If we now introduce the eigenvalues and eigenvectors of the Hamiltonian,
Eq.~(\ref{eqh19}) can be recast into:
\begin{equation}\label{eqh20}
{\mathcal Z}_0\langle x_1x_m\rangle = \lim_{N\rightarrow\infty}
\mbox{Tr} \left(\left[{\mathcal Q}\right]^{N-m} \left[{U}_{\mathcal
    Q}\right]^{\dagger} \left[{\mathcal Q}_X\right]\left[U_{\mathcal
    Q}\right]\left[{\mathcal Q}\right]^{m-2}\left[{U}_{\mathcal
    Q}\right]^{\dagger}\left[{\mathcal Q}_X\right]\left[U_{\mathcal
    Q}\right]\right).
\end{equation}

In this form we recognize the 
standard expression in Eq.~(\ref{eqh17a}). Indeed, 
the combination $\left[{U}_{\mathcal
    Q}\right]^{\dagger} \left[{\mathcal Q}_X\right]\left[U_{\mathcal
    Q}\right]$ represents the matrix elements of the hole density operator in
the basis of the eigenstates of the Hamiltonian, and the factors
$\left[{\mathcal Q}\right]$ the exponential factors in Eq.~(\ref{eqh17a}).  

\section{Spin 1/2 system}\label{Sec:spinful}

We turn now to the spinful case. In this section, we show how results
obtained for the spinless system are relevant and useful to derive in a
straightforward fashion the corresponding quantities of the spin 1/2 system. 

\subsection{Partition function}

Since up- and down-spins are decoupled, and since the fermionic contribution
to the action is identical for both of them, we can write the fermionic
determinant as: det$\left[S\right]$ = det$\left[S_0\right]^2$. This allows to
express the determinant as the product of the traces of the matrix products $\prod_{n=1}^N
\left[{\mathcal K}_n\right]$. By making use of the mixed product property of
the Kronecker product: 
$\left(\left[A\right]\left[C\right]\right)\otimes
\left(\left[B\right]\left[D\right]\right)= \left(\left[A\right]\otimes
\left[B\right]\right)\left(\left[C\right]\otimes
\left[D\right]\right)$, we obtain: 

\begin{equation}\label{eqh14}
\mbox{det} \left[S\right] = \mbox{Tr} \prod_{n=1}^N \left[{\mathcal
    K}_n\right] 
\otimes \left[{\mathcal K}_n\right].  
\end{equation}

The partition function is obtained by combining Eqs.~(\ref{eqh14}) and
(\ref{eqh2}). Since the time steps are decoupled, the evaluation of the
integrals over the $x$ and $\lambda$ fields is straightforward. The tensorial
products $\left[{\mathcal K}_n\right]  \otimes \left[{\mathcal K}_n\right]$
yield 16$\times$16 matrices, which, after application of the projectors
${\mathcal P}_1 \ldots {\mathcal P}_N$, become block diagonal with a $4
\times 4$ zero block. They all reduce to the same 12$\times$12 real
symmetric matrix  $\left[k\right]$. 
The matrix $\left[k\right]$ represents the projection of the tensor product
$\left[{\mathcal K}_n\right]\otimes \left[{\mathcal K}_n\right]$ for all $n$
as 

\begin{equation}\label{eqh14b}
\left[k\right]_{i,j} = {\mathcal P}_n (\left[{\mathcal K}_n\right]_{i_1,j_1}
\left[{\mathcal K}_n\right]_{i2,j2}) 
\end{equation}

\noindent with the convention 
\begin{equation}\label{convent}
i = \begin{cases} 4(i_1-1) + i_2 & \text{if ($1\leq i_1\leq 2$ and $1\leq
    i_2\leq 4$)}\\
& \quad \text{Êand ($i_1=3 $ and $1\leq i_2\leq 2$)},\\
i=i_2 +10 & \text{if ($i_1=4 $ and $1\leq i_2\leq 2$).} \end{cases}
\end{equation}
and similarly for $j$. The remaining
matrix elements of the $16\times 16$ matrix form a vanishing $4\times 4$ separate block
and have been discarded in Eq.~(\ref{eqh14b}).
Finally we obtain a simple expression for the partition function ${\mathcal Z}$ of the
two-site single impurity Anderson model:  

\begin{equation}\label{eqh15}
{\mathcal Z} = \lim_{N\rightarrow\infty} \mbox{Tr}~\left[k\right]^N.
\end{equation}
with
\begin{equation}\label{eqk}
\left[k\right]=
\left(\begin{array}{cccccccccccc}
1&&&&&&&&&&&\\
&L_{\rm c}& \delta V&&&&&&&&& \\
&\delta V& L_{\rm d}&&&&&&&&& \\
&&&L_{\rm c}L_{\rm d}&&&&&&&& \\
&&&&L_{\rm c}&0&0&0&\delta V&0&0&0 \\
&&&&0&L_{\rm c}^2&L_{\rm c}\delta V&0&0&L_{\rm c}\delta V&0&0 \\
&&&&0&L_{\rm c}\delta V&L_{\rm c}L_{\rm d}&0&0&0&0&0 \\
&&&&0&0&0&L_{\rm c}^2 L_{\rm d}&0&0&0&0 \\
&&&&\delta V&0&0&0&L_{\rm d}&0&0&0\\
&&&&0&L_{\rm c}\delta V&0&0&0&L_{\rm c}L_{\rm d}&0&0 \\
&&&&0&0&0&0&0&0&L_{\rm c} L_{\rm d}&0 \\
&&&&0&0&0&0&0&0&0&L_{\rm c}^2L_{\rm d}
\end{array}\right).
\end{equation}
\noindent When expanded to lowest order in $\delta$, the blocks of
$\left[k\right]$ represent the hamiltonian matrix in the Fock space, in the
same fashion as in Eq.~(\ref{eqh13b}). Diagonalizing these blocks yields the
expected 
expression of the partition function ${\mathcal Z}$:  

\begin{eqnarray}\label{eqh16}
{\mathcal Z} & = & 1 + 3 \exp \left[-\beta\left(\epsilon_{\rm c} +
  \epsilon_{\rm d} - 2\mu \right)\right] + 2
\exp\left[-\beta\left(2\epsilon_{\rm c} + \epsilon_{\rm d} -
  3\mu\right)\right] 
\nonumber \\
& &{}+ 2 \sum_{j=\pm 1}\exp\left[-\beta\left(\frac{\displaystyle \epsilon_{\rm
  c} + 
  \epsilon_{\rm d}}{\displaystyle 2} - \mu +j
  \sqrt{\left(\frac{\displaystyle \epsilon_{\rm c} - \epsilon_{\rm
  d}}{\displaystyle 2}\right)^2 + V^2}\right)\right] 
\nonumber \\
& &{} + \sum_{j=\pm 1}\exp\left[-\beta\left(\frac{\displaystyle 3\epsilon_{\rm
  c} + 
  \epsilon_{\rm d}}{\displaystyle 2} - 2\mu +j
  \sqrt{\left(\frac{\displaystyle \epsilon_{\rm c} - \epsilon_{\rm
  d}}{\displaystyle 2}\right)^2 + 2V^2}\right)\right] .
\end{eqnarray}

Through the exact calculation of the functional integrals we have recovered
this  result that can also be derived from the diagonalization of
the Hamiltonian. Note that $V$ is multiplied by different coefficients in
the one-particle and two-particle states. 
In contrast to the spinless case Eq.~(\ref{eqh13b}) the eigenvalues of the
hamiltonian matrix entering Eq.~(\ref{eqk}) result from entangled states, the
entanglement being achieved by the projection onto the physical Fock space in
Eq.~(\ref{eqh14b}). Note that we here obtained the hamiltonian matrix without
having explicitely used any basis of the Fock space. 
It naturally arose as the projected tensor product of the basis appropriate 
to the spinless case.

\subsection{Hole density and autocorrelation function}

In the spinful case, the hole density $\langle x_m\rangle$ is given by: 

\begin{eqnarray}\label{eqhx3}
{\mathcal Z}\langle x_m\rangle & = & \lim_{\substack{N\rightarrow\infty \\
  W\rightarrow\infty}} {\mathcal P_1}\ldots{\mathcal 
  P_N}~\left(\mbox{det} \left[S\right]~x_m\right)\\
  \nonumber
 & = & \lim_{\substack{N\rightarrow\infty \\ W\rightarrow\infty}} {\mathcal
  P_1}\ldots{\mathcal P_N}~\left(x_m \mbox{Tr}~\prod_{n=1}^N
\left[{\mathcal K}_n\right] \otimes \left[{\mathcal K}_n\right]\right) 
\end{eqnarray}

For the spin 1/2 case, the counterpart of the matrix $\left[{\mathcal
    Q}_X\right]$ is the matrix $\left[{\mathcal K}_X\right] \equiv {\mathcal
  P}_n (x_n \left[{\mathcal K}_n\right]\otimes \left[{\mathcal
    K}_n\right])$. It is a 12$\times$12 matrix, the elements of which may be
expressed as: 

\begin{equation}\label{eqhx3b}
\left[{\mathcal K}_X\right]_{i,j} = \delta_{i,1}~\delta_{j,1} +
\delta_{i,2}~\delta_{j,2}  + \delta_{i,5}~\delta_{j,5} +
\delta_{i,6}~\delta_{j,6}, 
\end{equation}

\noindent using Eqs.~(\ref{eqh4a})--(\ref{eqh4g}), in the $\delta \rightarrow
0$ limit. In this form it represents the hole density operator on
the impurity site in the Fock space. Thus Eq.~(\ref{eqhx3}) becomes

\begin{equation}\label{eqhx4}
{\mathcal Z}\langle x_m\rangle = \lim_{N\rightarrow\infty}
\mbox{Tr} \left(\left[{\mathcal K}_X\right]\left[k\right]^{N-1}\right),
\end{equation} 

\noindent which has exactly the same form as for the spinless case
and a similar interpretation applies.

As for autocorrelation functions such as $\langle x_1x_m\rangle$, we can adopt
the same procedure to obtain  

\begin{equation}\label{eqh19a}
{\mathcal Z}\langle x_1x_m\rangle = \lim_{N\rightarrow\infty}
\mbox{Tr} \left(\left[{\mathcal K}_X\right]\left[k\right]^{m-2}\left[{\mathcal
      K}_X\right]\left[k\right]^{N-m}\right),  
\end{equation} 

\noindent which, when again introducing the eigenstates of the Hamiltonian,
can also be identified to the ordinary expression in
Eq.~(\ref{eqh17a}). On the formal level, the use of the slave boson representation in
the radial gauge greatly simplifies the evaluation of the dynamical
correlation functions of the operators that can be represented by the radial
slave bosons. 

\section{Green's function}\label{Sec:GF}

We turn to the impurity Green's function $G_{\sigma}(q-p)$. When
expressed in terms of the eigenvalues of the Hamiltonian, $E_\alpha$ --- which
can be read from, e.~g., Eq.~(\ref{eqh16}) for spin 1/2 --- it reads: 

\begin{equation}\label{eqh34}
{\mathcal Z}G_{\sigma}(q-p) = -\sum_{\alpha,\alpha'} e^{\delta(q-N-p)
  E_{\alpha}} \langle 
\psi_{\alpha}|d_\sigma|\psi_{\alpha'}\rangle e^{\delta(p-q) E_{\alpha'}}
  \langle \psi_{\alpha'}|d^\dagger_\sigma|\psi_{\alpha}\rangle, 
\end{equation}

In the radial gauge the creation and annihilation operators are expressed in
terms of auxiliary fields as given in Eqs.~(\ref{eqh01a}) and
(\ref{eqh01b}). We obtain $G_{\sigma}(q-p)$ as follows:  

\begin{eqnarray}\label{eqh21}
{\mathcal Z}G_{\sigma}(q-p) = -\lim_{\substack{N\rightarrow\infty \\
    W\rightarrow\infty}}\prod_{n=1}^N \int && 
\prod_{\sigma'=\downarrow,\uparrow}
D[f_{n,\sigma'},f^\dagger_{n,\sigma'}]
D[c_{n,\sigma'},c^\dagger_{n,\sigma'}]\\
\nonumber &\times& \int_{-\infty}^{\infty}
\frac{\displaystyle \delta {\rm d}\lambda_n}{\displaystyle 2\pi}
\int_{-\infty}^{\infty} {\rm d}x_n~e^{-S}
f^{\phantom{\dagger}}_{q,\sigma}f^\dagger_{p,\sigma}x_{q+1} x_p, 
\end{eqnarray}

\noindent in the language of functional integrals. Note that the three other
Green's functions involving the three expectation values: $\langle
d^{\phantom{\dagger}}_{q,\sigma}c^\dagger_{p,\sigma}\rangle$, $\langle
c^{\phantom{\dagger}}_{q,\sigma}d^\dagger_{p,\sigma}\rangle$ and $\langle
c^{\phantom{\dagger}}_{q,\sigma}c^\dagger_{p,\sigma}\rangle$, can be
calculated in the same fashion, except that they are simpler to evaluate since
they contain at most one amplitude of the slave bosonic field, $x$, unlike the 
one we chose to study. 

\subsection{Derivation of the pseudofermion Green's function}

Following standard procedures (see, e.~g., Negele and Orland \cite{NEG88}), we
cast Eq.~(\ref{eqh21}) into the form: 

\begin{equation}\label{eqh22}
{\mathcal Z}G_{\sigma}(q-p) = -\lim_{\substack{N\rightarrow\infty \\
  W\rightarrow\infty}}{\mathcal P_1}\ldots{\mathcal 
  P_N}~\left(\mbox{det}\left[S_0\right] {\mathcal
  G}_{p,q}~x_px_{q+1}\right), 
\end{equation}

\noindent where ${\mathcal G}_{p,q}$ is the minor of one of the matrix 
elements of the
2$\times$2 block that shares the same row labels as $\left[{\mathcal
    L}_p\right]$ and column labels as $\left[{\mathcal L}_{q+1}\right]$ in the
matrix $\left[S_0\right]$ as defined in Eq.~(\ref{eqh6}) (if $q=N$ then the
block to be considered is $\left[{\mathcal L}_1\right]$). The minor ${\mathcal
  G}_{p,q}$ is the unprojected pseudofermion Green's function. 

For the subsequent calculations we set $p=N-m+1$ and $q=N$. The minor
${\mathcal G}_{N-m+1,N}$ can be calculated as:  

\begin{equation}\label{eqh23}
{\mathcal G}_{N-m+1,N} = \frac{\displaystyle \partial}{\displaystyle \partial
  a}~\mbox{det}\left[S(a;m)\right], 
\end{equation}

\noindent where $\left[S(a;m)\right]$ is given by:

\begin{equation}\label{eqh24}
\left[S(a;m)\right]_{i,j} = \left[S_0\right]_{i,j} +
a\delta_{i,2(N-m+1)}\delta_{j,2N}. 
\end{equation}

To calculate $\mbox{det}\left[S(a;m)\right]$, we find it convenient to move the
last two columns to the left in the matrix $\left[S(a;m)\right]$ in
Eq.~(\ref{eqh24}). In the same fashion as for $\mbox{det}\left[S_0\right]$, we
expand $\mbox{det}\left[S(a;m)\right]$ along the first two columns.  
Once the derivative of $\mbox{det}\left[S(a;m)\right]$ with
respect to $a$ is calculated we obtain the three
contributions:  

\begin{equation}\label{eqh25}
\frac{\displaystyle \partial}{\displaystyle \partial
  a}~\mbox{det}\left[S(a;m)\right] = -1\times 
{\mathcal M}_{1,a}^{\overline{1}} + 
L_{\rm c} {\mathcal M}_{L_{\rm c},a}^{\overline{1}} +
\delta Vx_1 {\mathcal M}_{\delta,a}^{\overline{1}}, 
\end{equation}

\noindent where the three minors ${\mathcal M}_{1,a}^{\overline{1}}$, 
${\mathcal M}_{L_{\rm c},a}^{\overline{1}}$ 
and ${\mathcal M}_{\delta,a}^{\overline{1}}$ are defined 
in the appendix. They can be
expressed as linear combinations of 
${\mathcal M}_{1,a}^{\overline{2}}$, $M_{L_{\rm c},a}^{\overline{2}}$ 
and ${\mathcal M}_{\delta,a}^{\overline{2}}$ in the same fashion
as for the minors defined in the previous sections. They also follow a
recurrence relation that reads:  

\begin{eqnarray}\label{eqh26}
\left(\begin{array}{c}
{\mathcal M}^{\overline{n}}_{1,a}\\
{\mathcal M}^{\overline{n}}_{L_{\rm c},a}\\
{\mathcal M}^{\overline{n}}_{\delta,a}
\end{array}\right)=
\left(\begin{array}{ccc}
1&0&0\\
0&L_{\rm c}&\delta V x_{n+1}\\
0&\delta V x_{n+1}&L_{n+1}
\end{array}\right)
\left(\begin{array}{c}
{\mathcal M}^{\overline{n+1}}_{1,a}\\
{\mathcal M}^{\overline{n+1}}_{L_{\rm c},a}\\
{\mathcal M}^{\overline{n+1}}_{\delta,a}
\end{array}\right),
\end{eqnarray}

\noindent up to time step $n=N-m$. Here we recognize the blocks defined in
Eqs.~(\ref{eqh9a}) and (\ref{eqh9b}) entering the matrix $[K^{(n+1)}]$ which we
encountered during the evaluation of the partition function. Then,
${\mathcal M}_{1,a}^{\overline{N-m}}$, 
${\mathcal M}_{L_{\rm c},a}^{\overline{N-m}}$ and
${\mathcal M}_{\delta,a}^{\overline{N-m}}$ are linear combinations of 
the minors ${\mathcal M}_{1,4}^{\overline{N-m+1}}$ 
and ${\mathcal M}_{3,4}^{\overline{N-m+1}}$:  

\begin{eqnarray}\label{eqh27}
\left(\begin{array}{c}
{\mathcal M}^{\overline{N-m}}_{1,a}\\
{\mathcal M}^{\overline{N-m}}_{L_{\rm c},a}\\
{\mathcal M}^{\overline{N-m}}_{\delta,a}
\end{array}\right)=
\left(\begin{array}{cc}
-1&0\\
0&L_{\rm c}\\
0&\delta V x_{N-m+1}
\end{array}\right)
\left(\begin{array}{c}
{\mathcal M}^{\overline{N-m+1}}_{1,4}\\
{\mathcal M}^{\overline{N-m+1}}_{3,4}
\end{array}\right).
\end{eqnarray}

The minors ${\mathcal M}_{1,4}^{\overline{n}}$ 
and ${\mathcal M}_{3,4}^{\overline{n}}$ are tightly related 
to $M_{1,4}^{\overline{n}}$ and $M_{3,4}^{\overline{n}}$: they are also built
on a matrix similar to $\left[S_0\right]$, but with the difference that only the
time steps $m$ with $m > n$ 
are included. ${\mathcal M}_{i,j}^{\overline{n}}$ is a minor of
this matrix, where the first two columns and both the $i$-th and $j$-th rows
are removed. Accordingly, the recurrence relations 
for ${\mathcal M}_{1,4}^{\overline{N-m+1}}$ 
and ${\mathcal M}_{3,4}^{\overline{N-m+1}}$ are also given in Eq.~(\ref{eqh9})
and hence we also need to introduce the minor
${\mathcal M}_{1,3}^{\overline{N-m+1}}$. Thus, their evaluation involves
the blocks of Eqs.~(\ref{eqh9b}) and (\ref{eqh9c}), which enter the
matrix $[K^{(n+1)}]$ jointly with the last three components of the column
vector in Eq.~(\ref{eqh10}), taken at time step 
$N$. Therefore, to calculate ${\mathcal G}_{N-m+1,N}$ as defined above we have
to consider the following set of six minors: ${\mathcal M}_{1,a}$, 
${\mathcal M}_{L_{\rm c},a}$, ${\mathcal M}_{\delta,a}$ for time 
steps $1\leq n \leq N-m$, and ${\mathcal M}_{1,3}$, ${\mathcal M}_{1,4}$
and ${\mathcal M}_{3,4}$ for time steps $N-m+2 \leq n \leq N$.  

Combining the above steps, we can write 
the unprojected pseudofermion Green's function as:

\begin{eqnarray}\label{eqh30}
{\mathcal G}_{N-m+1,N} = && \mbox{Tr} \left[
\left[K^{(N)}_f\right]
\times
\left(\prod_{n=N-1}^{N-m+2}\left[K^{(n)}_>\right]\right) 
\right.\\
\nonumber
&&~~~~\times \left.
\left[K^{(N-m+1)}_{f^{\dagger}}\right] 
\times
\left(\prod_{n=N-m}^{1}\left[K^{(n)}_<\right]\right) 
\right], 
\end{eqnarray}

\noindent where the time steps enter in decreasing order. The 6$\times$6 
matrices $\left[K^{(N)}_f\right]$
$\left(\left[K^{(N-m+1)}_{f^{\dagger}}\right]\right)$ 
representing the annihilation (creation) of a fermion in the world line
language are given by
\begin{eqnarray}
\left[K^{(N)}_f\right]_{i,j} &=& \delta V x_{N} ~\delta_{i,1}~\delta_{j,4} +
L_{N}~\delta_{i,2}~\delta_{j,5} + L_{\rm c}
L_N~\delta_{i,2}~\delta_{j,6}\nonumber\\
\left[K^{(N-m+1)}_{f^{\dagger}}\right]_{i,j} &=& 1~\delta_{i,5}~\delta_{j,1} +
L_{\rm c}~\delta_{i,6}~\delta_{j,2} + 
\delta V x_{N-m+1}~\delta_{i,6}~\delta_{j,3} ~.
\end{eqnarray}
In Eq.~(\ref{eqh30}), $\left[K^{(n)}_<\right]$ is a 6 $\times$ 6 block
diagonal matrix the non zero elements of which are the two blocks
$\left[K^{(n),1}\right]$ and $\left[K^{(n),2}\right]$ defined in
Eqs.~(\ref{eqh9a}) and (\ref{eqh9b}). The matrix $\left[K^{(n)}_>\right]$ is
also 6 $\times$ 6 block diagonal matrix determined by 

\begin{equation}\label{eqksup}
\left[K^{(n)}_>\right]=\left[K^{(n)}\right] - \left[K^{(n)}_<\right].
\end{equation}

Equation (\ref{eqh30}) can also be interpreted on the basis of
Fig.~\ref{fig1}. In fact, computing 
$\langle d^{\phantom{\dagger}}_{N,\sigma}d^\dagger_{N-m+1,\sigma}\rangle$ 
can be visualized as the resummation of particular subsets of world
lines. They are 
naturally split into sets involving $L_{\rm d}$ for $N-m+2 < n < N-1$, in
which case $\left[K^{(n)}_>\right]$ is controlling the dynamics of the world
lines; and sets excluding $L_{\rm d}$ for $1 < n < N-m$, in which case
$\left[K^{(n)}_<\right]$ controls the dynamics. The transition from the
first (second) subset to the second (first) is taken care of by the matrix
$\left[K^{(N-m+1)}_{f^{\dagger}}\right]$
$\left(\left[K^{(N)}_f\right]\right)$. 

This expression for the Green's function in the world line language can also
be related to its counterpart in the hamiltonian
language. Indeed, the trace of the matrix product in Eq.~(\ref{eqh30}) can be
written in terms of 4$\times$4 matrices:  

\begin{equation}\label{eqh30a}
{\mathcal G}_{N-m+1,N} = \mbox{Tr} 
\left(
\left[F_N\right] \times 
\left( \prod_{n=N-1}^{N-m+2}\left[{\mathcal K}^>_n\right] \right)\times
\left[\Phi_{N-m+1}\right] \times
\left( \prod_{n=N-m}^{1}\left[{\mathcal K}^<_n\right] \right)
\right), 
\end{equation}

\noindent where

\begin{equation}\label{eqh29}
\left[\Phi_{N-m+1}\right] = \left(\begin{array}{cccc}
0&0&0&0\\
0&0&0&0\\
1&0&0&0\\
0&L_{\rm c}&\delta V x_{N-m+1}&0
\end{array}\right)~~~~ \mbox{and} ~~~~ 
\left[F_N\right] = \left(\begin{array}{cccc}
0&\delta V x_N&L_{N}&0\\
0&0&0&L_{\rm c}L_N\\
0&0&0&0\\
0&0&0&0
\end{array}\right),
\end{equation}

\noindent characterize the creation and annihilation of 
an electron, respectively. The matrices $\left[{\mathcal K}^<_n\right]$ and $\left[{\mathcal
    K}^>_n\right]$ result from $\left[{\mathcal K}_n\right]$ as: 
    
\begin{eqnarray}\label{eqKinf}
\left[{\mathcal K}^<_n\right]_{i,j} &=& \left[{\mathcal K}^n\right]_{i,j} -
L_{\rm c} L_{n} ~\delta_{i,4}~\delta_{j,4}\; , \nonumber\\ 
\left[{\mathcal K}^>_n\right]_{i,j} &=& \left[{\mathcal K}^n\right]_{i,j} -
\delta_{i,1}~\delta_{j,1}~. 
\end{eqnarray}

\subsection{Spinless case}

Now that the unprojected pseudofermion Green's function 
has been converted to a compact form, we can evaluate the
physical Green's function which, in the spinless case, reads: 

\begin{eqnarray}\label{eqg0}
\nonumber
{\mathcal Z}_0G_0(q-p) &=& -\lim_{\substack{N\rightarrow\infty \\
    W\rightarrow\infty}} {\mathcal P_1}\ldots{\mathcal P_N}~ 
\mbox{Tr}\left[ x_1x_{N-m+1}
\left[F_N\right]
\times 
\left(\prod_{n=N-1}^{N-m+2}\left[{\mathcal K}^>_n\right]\right) 
\right.\\
&\times&\left.
\left[\Phi_{N-m+1}\right]
\times 
\left(\prod_{n=N-m}^{1}\left[{\mathcal K}^<_n\right]\right)
\right].
\end{eqnarray}

With the application of the projectors ${\mathcal P_1}\ldots{\mathcal P_N}$, we
obtain: 
 
\begin{equation}\label{eqg1}
{\mathcal Z}_0G_0(q-p) = -\lim_{N\rightarrow\infty} \mbox{Tr} \left(
\left[{\mathcal Q}_X\right] \left[{\mathcal F}\right] 
 \left[q^>\right]^{m-2}\left[\phi\right]
\left[q^<\right]^{N-m-1} \right),
\end{equation}

\noindent where $\left[{\mathcal Q}_X\right]$ is given by
Eq.~(\ref{eqdefqx}). The matrices $\left[\phi\right]$ and 
$\left[{\mathcal F}\right]$ are defined as $\left[\phi\right] \equiv {\mathcal
  P}_{N-m+1} \left(x_{N-m+1}\left[\Phi_{N-m+1}\right]\right)$ and
$\left[{\mathcal F}\right] \equiv {\mathcal P}_N
\left(\left[F_N\right]\right)$ respectively. They can be read off from
Eq.~(\ref{eqh29}) if $x$ is replaced by 1, and $L_{n}$ by $L_{\rm d}$. 
Note that $\left[\phi\right] = \left[{\mathcal F}\right]^\dagger$ only in the
limit $\delta \rightarrow 0$ when $L_{\rm c} \rightarrow 1$ and $L_{\rm d}
\rightarrow 1$. In the limit $\delta \rightarrow 0$, they coincide with the
matrix representations of the operators $f^{\dagger}$ and $f$
respectively. The matrices $\left[q^<\right]$ and $\left[q^>\right]$ are given
by $\left[q^<\right]\equiv {\mathcal P}_n (\left[{\mathcal K}^<_n\right])$ and
$\left[q^>\right]\equiv {\mathcal P}_n (\left[{\mathcal K}^>_n\right])$
respectively, and are easily related to $\left[q\right]$:
\begin{eqnarray}\label{eqqpp}
\left[q^<\right]_{i,j} & = & \left[q\right]_{i,j} - L_{\rm c}L_{\rm
  d}~\delta_{i,4}~\delta_{j,4} \nonumber\\
\left[q^>\right]_{i,j} & = & \left[q\right]_{i,j} - ~\delta_{i,1}~\delta_{j,1}.
\end{eqnarray}

we can now reshape Eq.~(\ref{eqg1}) into the form: 

\begin{eqnarray}\label{eqg1c}
{\mathcal Z}_0G_0(q-p) & = &-\lim_{N\rightarrow\infty} \mbox{Tr} \left[
\left[{\mathcal Q}^<\right]^{N-m-1}
\left(\left[U_{\mathcal Q}\right]^{\dagger}
\left[{\mathcal Q}_X\right]\left[{\mathcal F}\right]
\left[U_{\mathcal Q}\right]\right)\right.
\\
\nonumber
&&~~~~~~~~~~~~~~\times \left.
\left[{\mathcal Q}^>\right]^{m-2}
\left(\left[U_{\mathcal Q}\right]^{\dagger}
\left[\phi\right]
\left[U_{\mathcal Q}\right]\right)
\right],
\end{eqnarray}

\noindent where $\left[{\mathcal Q}^<\right]$ and $\left[{\mathcal
    Q}^>\right]$ are obtained from the diagonalization of $\left[q^<\right]$
and $\left[q^>\right]$, respectively. Note that the same  matrix
$\left[U_{\mathcal Q}\right]$ diagonalizes the matrices
$\left[q\right],~\left[q^<\right]$, and $\left[q^>\right]$. 

In Eq.~(\ref{eqg1c}), the product 
$\left[U_{\mathcal Q}\right]^{\dagger}
\left[{\mathcal Q}_X\right] \left[{\mathcal F}\right]
\left[U_{\mathcal Q}\right]$ is the representation of
the annihilation operator $d$ in the basis of the eigenstates of the
Hamiltonian, while $\left[U_{\mathcal
    Q}\right]^{\dagger}\left[\phi\right]\left[U_{\mathcal Q}\right]$
represents $d^\dagger$, as can be easily verified explicitly in the limit
$\delta \rightarrow 0$. The factors $\left[{\mathcal Q}^>\right]$ and
$\left[{\mathcal Q}^<\right]$ determine the time evolution.  
Therefore Eq.~(\ref{eqg1c}) can be easily identified with
Eq.~(\ref{eqh34}). Note that all eigenvalues seem to contribute to the
Green's function in Eq.~(\ref{eqh34}), and only the matrix elements of the
creation (annihilation) operators restrict the set of eigenvalues that
effectively contribute to the Green's function. 
In contrast, in
Eq.~(\ref{eqg1c}) some of these restrictions are contained in the factors
$\left[{\mathcal Q}^>\right]$ and $\left[{\mathcal Q}^<\right]$ which
replace the full set of eigenvalues,  that would be contained in the
matrix $\left[{\mathcal Q}\right]$. 

It is tempting to compare the physical electron Green's function (including the
factors $x$) to the projected pseudo-fermion Green's function (without the
factors $x$). Straightforward algebra yields: 

\begin{equation}\label{eqhcomp1}
\lim_{\delta\rightarrow 0}~{\mathcal
  P}_{N-m+1}\left(x_{N-m+1}\left[\Phi_{N-m+1}\right]\right) =  
\lim_{\delta\rightarrow 0}~{\mathcal
  P}_{N-m+1}\left(\left[\Phi_{N-m+1}\right]\right), 
\end{equation}

\noindent and

\begin{equation}\label{eqhcomp2}
\lim_{\delta\rightarrow 0}~{\mathcal P}_1{\mathcal P}_N
\left( x_1 \left[K_1^<\right] \left[F_N\right]\right) =  
\lim_{\delta\rightarrow 0}~{\mathcal P}_1{\mathcal P}_N \left(
  \left[K_1^<\right] \left[F_N\right]\right). 
\end{equation}

Therefore, \emph{as a particularity of the spinless case}, the factors $x$ in
Eq.~(\ref{eqg0}) play no role, and both Green's functions
coincide. Consequently the same result for the physical electron Green's
function would have been obtained by substituting $x$ by 1 in
Eqs.~(\ref{eqh01a}) and 
(\ref{eqh01b}), and accordingly in the fermionic contribution to the action
$S_f$. Incidentally, such a procedure is in complete agreement with the
original suggestion by Kotliar and Ruckenstein \cite{KOT86} to modify the
expression of the physical electron operator by introducing square root
factors, when extended to the spinless case. 

\subsection{Spin 1/2 system}

Again, as shown below, results obtained for the spinless system can be
immediately applied to derive the Green's function in the spinful
case. Inserting Eq.~(\ref{eqh30}) into Eq.~(\ref{eqh22}) yields: 

\begin{eqnarray}\label{eqh31}
{\mathcal Z}G_{\sigma}(q-p) &=& -\lim_{\substack{N\rightarrow\infty \\
  W\rightarrow\infty}} {\mathcal P_1}\ldots{\mathcal 
  P_N}~x_1x_{N-m+1}\nonumber\\
&\mbox{Tr}&\left[
\left(\left[{\mathcal K}_N\right]\otimes\left[F_N\right]\right) \times
\left(\prod_{n=N-1}^{N-m+2}\left[{\mathcal
    K}_n\right]\otimes\left[{\mathcal K}^>_n\right]\right)
\right.\\
\nonumber
&\times& \left.
\left(\left[{\mathcal
    K}_{N-m+1}\right]\otimes\left[\Phi_{N-m+1}\right]\right)
\times 
\left(\prod_{n=N-m}^{1}\left[{\mathcal
    K}_n\right]\otimes\left[{\mathcal K}^<_n\right]\right) 
\right], 
\end{eqnarray}

\noindent which after application of the projectors 
${\mathcal P_1}\ldots{\mathcal P_N}$ becomes:

\begin{equation}\label{eqh32}
{\mathcal Z}G_{\sigma}(q-p) = -\lim_{N\rightarrow\infty} \mbox{Tr}
\left(\left[{\mathcal K}_X\right] \left[\xi\right]
\left[k^>\right]^{m-2}
\left[\varphi \right] \left[k^<\right]^{N-m-1}
\right),
\end{equation}

\noindent with the matrices $\left[\varphi\right] = {\mathcal P}_{N-m+1} 
\left(x_{N-m+1}\left[{\mathcal
      K}_{N-m+1}\right]\otimes\left[\Phi_{N-m+1}\right]\right)$ and 
$\left[\xi\right] = {\mathcal P}_N \left(\left[{\mathcal
      K}_N\right]\otimes\left[F_N\right]\right)$. 
Leaving out entries which do not contribute in the limit $N\rightarrow\infty$
they read:
\begin{eqnarray}\label{eqvarphixi}
\left[\varphi\right]_{i,j} &=& 1 ~\delta_{i,3} ~\delta_{j,1} 
+ L_{\rm c} (\delta_{i,4} ~\delta_{j,2} +\delta_{i,7} ~\delta_{j,5} )
+ L_{\rm c}^2 ~\delta_{i,8} ~\delta_{j,6} \nonumber\\
\left[\xi\right]_{i,j} &=&   L_{\rm c} ~\delta_{i,1} ~\delta_{j,3} 
+ L_{\rm c}  L_{\rm d} (\delta_{i,2} ~\delta_{j,4}+\delta_{i,5} ~\delta_{j,7})
+ L_{\rm c}^2  L_{\rm d} ~\delta_{i,6} ~\delta_{j,8}~.
\end{eqnarray}
The
matrices $\left[k^<\right]$ and $\left[k^>\right]$ are given by
$\left[k^<\right]\equiv {\mathcal P}_n (\left[{\mathcal
    K}_n\right]\otimes\left[{\mathcal K}^<_n\right])$ and
$\left[k^>\right]\equiv {\mathcal P}_n (\left[{\mathcal
    K}_n\right]\otimes\left[{\mathcal K}^>_n\right])$ respectively. The
asymmetry in the representation of the physical electron creation and
annihilation operators (Eqs.~(\ref{eqh01a}) and (\ref{eqh01b})) is also
apparent in Eq.~(\ref{eqh32}). Indeed the operator $d^\dagger_{\sigma}$ is
represented by the matrix $\left[\varphi\right]$, and 
$d^{\phantom{\dagger}}_{\sigma}$ by the product of the matrices
$\left[{\mathcal K}_X\right]\left[\xi\right]$. In contrast to the spinless
case the factors $x$ in Eq.~(\ref{eqh31}) play a role, and the projected
pseudofermion Green's function and  physical electron Green's function
differ. 

\section{Conclusion}

In summary we have established a new scheme which provides a fundamental
connection between the 
representation of expectation values and dynamical correlation functions in
the hamiltonian language and their counterpart in the slave-boson path
integral formulation. This has been achieved for the $U=\infty$ spin-1/2
single impurity Anderson model through their exact evaluation for a two site
cluster. The new scheme allowed us to compute the partition function and the
hole density, expressed as the expectation value of the radial slave field
$x$. Moreover the Green's function and the hole auto-correlation function 
were evaluated within this scheme.  

We verified that
the exact expectation value of the slave boson amplitude field $x$ is finite,
as postulated in mean-field calculations, even in this extreme quantum
case. It is therefore not related to the condensation of a boson, which would
necessarily vanish in such a calculation. 
The suppression of the condensation 
originates in the use of the
radial representation, where the phase of the boson is integrated out in the
first place. We note that higher slave boson correlation functions
such as $\langle x^n(\tau) x^m(0) \rangle $ reduce to  $\langle x(\tau) x(0)
\rangle $. Therefore the field $x$ bears little resemblance 
to ordinary
complex bosonic fields. The corresponding calculations follow a similar scheme
as those for the partition function.

Through an independent calculation we obtained both the physical electron 
and pseudofermion Green's functions. In the spinless case, the projected
pseudofermion Green's function is finite, and it is identical to the one of
the physical electron. Therefore a ``perturbation theory''-like
factorization of the latter as a product of the boson and pseudofermion
Green's functions does not appear appropriate in general, but it may still
be valid in particular frequency ranges, such as the low frequency domain.
In the latter case, a mean-field decoupling looks more appropriate. 
It is likely to provide a better agreement with the exact result if the square
root factors, originally proposed in \cite{KOT86}, are introduced. 

It is also of great importance to understand that our formalism allows
immediate and straightforward use of results, which were obtained for the
spinless system, in order to derive those of the spinful case: 
the proposed scheme first treats the coherent states of fermions in the two 
spin sectors (up and down spin) separately. The world lines of particles with
different spin projection evolve independently. Only in a final step, when the 
full fermionic determinant is built from the product of the determinants of
the  two spin species, the projection onto the physical space with no double
occupancies 
is straightforwardly achieved through the projection rules, 
Eqs.~(\ref{eqh4a})--(\ref{eqh4g}), 
applied to the entries of the determinant. Here, the projection is easily 
accomplished in the Fock space which, 
when directly enforced for the world lines, 
produces a  complicated entanglement of coherent states.
For larger systems the exact resummation of the world lines and their
respective projection --- as presented in, e.g., Eqs.~(\ref{eqh19}),
(\ref{eqh19a}) and (\ref{eqg1}) --- is difficult on the analytical level, but
probably not on the numerical level. As an alternative to the exact
calculation  one may consider a plain saddle point approximation scheme when
tackling spatially extended systems. Unfortunately the latter fails to
reproduce 
the exact result even for the two-site problem, and it will be necessary to
determine appropriate quasiparticle weight factors from the two-site solution
within an effective slave boson approach, similar to the Kotliar-Ruckenstein
scheme \cite{KOT86}. This challenge will be addressed in future work
\cite{Fre07}. 

The extension of the above scheme to the spin rotation invariant
formulation of the $t$-$J$ model where the phases of all the bosonic fields
can be gauged away is desirable \cite{FRE92}. Work along this line is in
progress.

One may wish to extend such a calculation to other representations of this
model, such as the ones based on Hubbard X-operators \cite{Tue95}. This
unfortunately poses another challenge, since the ``angular part'' of the
respective action is intrinsically off diagonal in time, which makes the
integral over the angular variables significantly more difficult. This also
holds true for the Kotliar-Ruckenstein representation where one of the bosonic
fields is complex \cite{FRE92,KOT86,JOL91}. Alternatively one may also
consider weak-coupling approaches, such as the Hubbard-Stratanovich decoupling
of the interaction term in the charge channel. Even if it were possible to
evaluate the partition function exactly, using the corresponding form of
Eqs.~(\ref{eqh2}), (\ref{eqh3}) and (\ref{eqh14}), it would still require a
major effort to obtain dynamical response functions. Nevertheless such a
calculation deserves further study.  

\section*{Acknowledgments}
We gratefully thank  O. Juillet for stimulating discussions. R.~F.\ is
grateful for the warm hospitality at the EKM of Augsburg University
where part of this work has been done. This work was
supported by the Deutsche Forschungsgemeinschaft (DFG) through SFB~484.
\appendix

\section{Expression of the minors ${\mathcal M}_{1,a}^{\overline{1}}$, 
${\mathcal M}_{L_{\rm c},a}^{\overline{1}}$ and 
${\mathcal M}_{\delta,a}^{\overline{1}}$} 

The minors ${\mathcal M}_{1,a}^{\overline{1}}$, 
${\mathcal M}_{L_{\rm c},a}^{\overline{1}}$ and 
${\mathcal M}_{\delta,a}^{\overline{1}}$ in Eq.~(\ref{eqh25}) are obtained 
from the expansion of $\mbox{det}\left[S(a;m)\right]$ given 
below for $p=N-m+1$. Their definition follows as:
\begin{itemize}
  \item ${\mathcal M}_{1,a}^{\overline{1}}$ by expanding 
$\mbox{det}\left[S(a;m)\right]$ along the two first columns and 
eliminating the $(2 N-1)$th and $2p$th lines.
  \item ${\mathcal M}_{\delta,a}^{\overline{1}}$ by expanding 
$\mbox{det}\left[S(a;m)\right]$ along the two first columns and 
eliminating the second and $2p$th lines.
  \item ${\mathcal M}_{L_{\rm c},a}^{\overline{1}}$ by expanding 
$\mbox{det}\left[S(a;m)\right]$ along the two first columns and eliminating 
the first and $2p$th lines.
\end{itemize}
In the expansion of $\mbox{det}\left[S(a;m)\right]$ 
${\mathcal M}_{1,a}^{\overline{1}}$ 
is multiplied by $\left[S(a;m)\right]_{2N-1,1} = 1$ and
$-\left[S(a;m)\right]_{2p,2} = -a$, ${\mathcal M}_{\delta,a}^{\overline{1}}$ is
multiplied by $-\left[S(a;m)\right]_{2,1} = \delta Vx_1$ and
$\left[S(a;m)\right]_{2p,2} = a$, 
and ${\mathcal M}_{L_{\rm c},a}^{\overline{1}}$ is multiplied 
by $\left[S(a;m)\right]_{1,1} = L_{\rm c}$ 
and $\left[S(a;m)\right]_{2p,2} = a$. The three minors above satisfy a
recurrence relation given in Eqs.~(\ref{eqh26}) and (\ref{eqh27}), while
$\mbox{det}\left[S(a;m)\right]$ reads:  

\begin{eqnarray}
&&\mbox{det}\left[S(a;m)\right] =  \\
&&\left|\begin{array}{cccccccccccccc}
L_{\rm c}&-\delta Vx_1&1&0&&&&&&&&&\\
-\delta Vx_1&L_1&0&1&&&&&&&&&&\\
&&-L_{\rm c}&\delta Vx_2&&&&&&&&&&\\
&&\delta Vx_2&-L_2&&&&&&&&&&\\
&&&&&&&&&&&&&\\
&&&&\ddots&\ddots&&&&&&&&\\
&&&&&&&&&&&&&\\
0&0&\ldots&&&&-L_{\rm c}&\delta Vx_p&1&0&&&&\\
0&a&\ldots&&&&\delta Vx_p&-L_p&0&1&&\\
&&&&&&&&-L_{\rm c}&\delta Vx_{p+1}&&&\\
&&&&&&&&\delta Vx_{p+1}&-L_{p+1}&&&&\\
&&&&&&&&&&&&&\\
&&&&&&&&&\ddots&\ddots&&&\\[-4em]
&&&&&&&&&&&&1&0\\
&&&&&&&&&&&&0&1\\
1&0&&&&&&&&&&&-L_{\rm c}&\delta Vx_N\\
0&1&&&&&&&&&&&\delta Vx_N&-L_N
\end{array}\right|
\nonumber
\end{eqnarray}


\begin{thebibliography}{999}
\bibitem{Pai00} S. Pairault, D. S\'en\'echal, A.-M.~S. Tremblay,
  Eur. Phys. J. B 16 (2000) 85; 
Phys. Rev. Lett. 80 (1998)  5389.
\bibitem{Anderson70} P.W. Anderson, J. Phys. C -- Solid State Phys. 3 (1970) 2346.
\bibitem{Nozieres74} P. Nozi\`eres, J. Low. Temp. Phys. 17 (1974) 31.
\bibitem{Wilson75} K.G. Wilson, Rev. Mod. Phys. 47 (1975) 773.
\bibitem{Andrei80} N. Andrei, Phys. Rev. Lett. 45 (1980) 379.
\bibitem{Wiegmann80} P.B. Wiegmann, Sov. Phys. JETP Lett. 31
(1980) 392.
\bibitem{Georges92} 
A. Georges, G. Kotliar, Phys. Rev. B 45 (1992) 6479.
\bibitem{Georges96} 
A. Georges, G. Kotliar, W. Krauth, M. J. Rozenberg,
Rev. Mod. Phys. 68 (1996) 13.
\bibitem{Metzner89} W. Metzner, D. Vollhardt, 
Phys. Rev. Lett. 62 (1989) 324.
\bibitem{Maier05} T. Maier, M. Jarrell, Th. Pruschke, M. H. Hettler, 
Rev. Mod. Phys. 77 (2005) 1027.
\bibitem{cdmft} G. Kotliar, S.Y. Savrasov, G. Palsson, G. Biroli, Phys.
Rev. Lett. 87 (2001) 186401.
\bibitem{bk} G. Biroli, G. Kotliar, Phys. Rev. B 65 (2002) 155112.
\bibitem{Potthoff05} M. Potthoff, Adv. Solid State Phys. 45 (2005) 135.
\bibitem{Kyung06} B. Kyung, S. S. Kancharla, D. S\'en\'echal,
 A.-M.~S. Tremblay, M. Civelli, G. Kotliar, Phys. Rev. B 73 (2006) 165114.
\bibitem{BAR76} S. E. Barnes, J. Phys. F: Metal Phys. 6 (1976) 1375.
\bibitem{Col84} P. Coleman, Phys. Rev. B 29 (1984) 3035. 
\bibitem{KOT86} G. Kotliar, A. E. Ruckenstein, Phys. Rev. Lett. 57 (1986)
  1362. 
\bibitem{FRE92}R. Fr\'esard, P. W\"olfle,
Int. J. Mod. Phys. B 6 (1992) 685; Erratum,
Int. J. Mod. Phys. B 6 (1992) 3087.
\bibitem{Has97} R. Fr\'esard, G. Kotliar,  Phys. Rev. B 56 (1997) 12909; 
                H. Hasegawa, {\it ibid.} 56 (1997) 1196.
\bibitem{Fre91}  R.~Fr\'esard, M.~Dzierzawa, 
P.~W\"olfle, Europhys. Lett. 15 (1991) 325.
\bibitem{Zimmermann97} W.~Zimmermann, R. Fr\'esard, P. W\"olfle,
  Phys. Rev. B 56 (1997) 10097. 
\bibitem{Flo02} S. Florens, A. Georges, G. Kotliar, O. Parcollet,
     Phys. Rev. B 66 (2002) 205102. 
\bibitem{Elitzur75} S. Elitzur, Phys. Rev. D 12 (1975) 3978. 
\bibitem{Kroha97} J. Kroha, P. W\"olfle, T.A. Costi,
Phys. Rev. Lett. 79 (1997) 261.
\bibitem{Kirchner04} S. Kirchner, J. Kroha,  P. W\"olfle,
Phys. Rev. B 70 (2004) 165102.
\bibitem{Baumgartner05} K. Baumgartner,  H. Keiter, phys stat sol (b) 242
  (2005)  377. 
\bibitem{Read} N. Read,  D. M. Newns, J. Phys. C 16 (1983) 3273. 
\bibitem{FRE01} R. Fr\'esard,  T. Kopp, Nucl. Phys. B 594 (2001) 769.
\bibitem{OUE07} H. Ouerdane, R. Fr\'esard,  T. Kopp, unpublished.
\bibitem{NEG88} J. W. Negele,  H. Orland, Quantum Many-Particle
  Systems (Addison-Wesley, 1988). 
\bibitem{BAR77} S. E. Barnes, J. Phys. F: Metal Phys. 7 (1976) 2637.
\bibitem{Tue95} E. O. T\"ungler,  T. Kopp, Nucl. Phys. B 443 (1995) 516.
\bibitem{JOL91} Th. Jolic{\oe}ur,  J. C. Le Guillou, Phys. Rev. B 44 (1991)
  2403. 
\bibitem{Fre07} R. Fr\'esard et al., unpublished.
\end{thebibliography}
\end{document}